\documentclass[aps,prev,twocolumn,preprintnumbers,floatfix,nofootinbib]{revtex4-1}

\usepackage{graphicx}
\usepackage{bm}
\usepackage{times}
\usepackage{slashed}
\usepackage{color}
\usepackage{epsfig}\usepackage{dcolumn}
\usepackage{color}
\def\br{\begin{eqnarray}}
\def\er{\end{eqnarray}}
\def\be{\begin{equation}}
\def\ee{\end{equation}}

\newcommand{\mzp}{M_{Z'}}
\newcommand{\mchi}{M_{N_1}}

\begin{document}

\title{Lepton Flavor Violation Induced by Dark Matter}

\author{Giorgio Arcadi$^1$}
\author{C. P. Ferreira$^2$}
\author{Florian Goertz$^1$}
\author{M. M. Guzzo$^2$}
\author{Farinaldo S. Queiroz$^{1,3}$}
\author{A.C.O. Santos$^{4,5}$}

\email{arcadi@mpi-hd.mpg.de}
\email{ferreira@ifi.unicamp.br}
\email{guzzo@ifi.unicamp.br}
\email{goertz@mpi-hd.mpg.de}
\email{farinaldo.queiroz@iip.ufrn.br}
\email{antonio_santos@fisica.ufpb.br}

\affiliation{$^1$Max-Planck-Institut f\"ur Kernphysik, Saupfercheckweg 1, 69117 Heidelberg, Germany\\
$^2$Instituto de F\'isica Gleb Wataghin - UNICAMP, 13083-859, Campinas SP, Brazil\\
$^3$International Institute of Physics, Federal University of Rio Grande do Norte, Campus Universit\'ario, Lagoa Nova, Natal-RN 59078-970, Brazil\\
$^4$Centre for Cosmology, Particle Physics and Phenomenology (CP3), Universit catholique de Louvain,B-1348, Louvain-la-Neuve, Belgium\\
$^5$Departamento de F\'isica, Universidade Federal da Para\'iba, Jo\~ao Pessoa-PB, 58051-970, Brazil\\
}

\begin{abstract}
Guided by gauge principles we discuss a predictive and falsifiable UV complete model where the Dirac fermion that accounts for the cold dark matter abundance in our universe induces the lepton flavor violation (LFV) decays $\mu \rightarrow e\gamma$ and $\mu \rightarrow e e e$ as well as $\mu-e$ conversion. We explore the interplay between direct dark matter detection, relic density, collider probes and lepton flavor violation to conclusively show that one may have a viable dark matter candidate yielding flavor violation signatures expected to be fully probed in the upcoming of experiments. Interestingly, keeping the dark matter mass not far from the TeV, our model has an approximate prediction for the maximum LFV signal one could have while reproducing the correct dark matter relic density. 
\end{abstract} 
    
\maketitle

\section{Introduction}
\label{introduction}

The fundamental particle nature of the dark matter is one of the most pressing questions in science. Therefore there is an intense search for signals from dark matter particles that could unveil its nature \cite{Bertone:2010zza,Bertone:2016nfn}. Among the dark matter candidates in the literature, WIMPs (weakly interacting massive particles) stand out for being able to elegantly yield the right dark matter relic density, as indicated by Planck \cite{Griest:1988yr,Griest:1988ma,Ade:2015xua}, in well-motivated theoretical models, and for predicting signals at ongoing and near future experiments \cite{Arcadi:2017kky}.

These experimental searches for WIMPs are classified into three categories: direct \cite{Akerib:2016vxi,Hehn:2016nll,Cui:2017nnn,Aprile:2017iyp,Amole:2017dex}, indirect \cite{Ade:2015xua,Abdallah:2016ygi,Abeysekara:2017jxs,Acharya:2017ttl} and collider probes \cite{Goodman:2010yf,Goodman:2010ku,Abercrombie:2015wmb}. Direct detection refers to the measurement of the WIMP-nucleon scattering rate at underground detectors \cite{Undagoitia:2015gya}. Indirect detection relies on the measured flux of cosmic-rays, gamma-rays and neutrinos that might feature excesses above astrophysical expectations \cite{Bringmann:2012ez,Profumo:2013yn}. As for colliders, WIMPs constitute simply events with a large missing transverse energy alongside visible counterparts \cite{Kahlhoefer:2017dnp}.  

As an attempt to map the allowed interactions between WIMPs and the Standard Model  (SM) particles, simplified models became powerful tools. A more appealing approach would be, on the other hand, represented by connecting the solution of the DM puzzle to other phenomena,for example with neutrino masses \cite{Dias:2010vt,Restrepo:2013aga,Queiroz:2014yna,Vicente:2014wga,Rodejohann:2015lca,Patra:2015vmp,Lindner:2016kqk,Gu:2016ghu,Ibarra:2016dlb,Dev:2016qeb,Kownacki:2017uyq,Ma:2017xxj}.

Following this philosophy, we discuss, in this work, the possibility of generating lepton flavor violation (LFV) via dark matter in a UV complete model. Lepton flavor violation is a clear signal of new physics.  In the SM, lepton flavor is a conserved
quantity since neutrinos are massless. Albeit, we have experimental confirmation that neutrinos are massive and experience flavor oscillations. Consequently, neutrino oscillations constitute a clear proof that lepton flavor is not a quantity conserved by Nature. However, we have not observed LFV between charged leptons yet. In this work, we explore the connection between dark matter and LFV, in particular, the processes $\mu \rightarrow e\gamma$, $\mu \rightarrow eee$ and $\mu-e$ conversion. 

In order to tightly connect LFV and dark matter via gauge principles, we arrange charged leptons and dark matter in the same multiplet of $SU(3)$. In this case, the $SU(3)$ triplet is comprised of an electron, electron neutrino, and a neutral fermion later to be identified as dark matter. This multiplet structure is replicated among the three generations to embed the SM fermionic content. Therefore, we have three neutral Dirac fermions, the lightest one being a dark matter candidate. Because of this extended gauge sector, new gauge bosons arise. One of them is a massive singly charged gauge boson, a $W^\prime$ which has been extensively searched for at the Large Hadron Collider (LHC) \cite{Jezo:2014wra,Chatrchyan:2014koa,Khachatryan:2016jww,Sirunyan:2017ukk}. This new boson connects, through charged current interactions, the neutral Dirac fermions, and the SM charged leptons. LFV processes occur then via the $W^\prime$ exchange involving such Dirac fermions. Since the lightest Dirac fermion is our dark matter, LFV decays are induced by dark matter. Our reasoning is based on the model proposed in \cite{Mizukoshi:2010ky}, which required the presence of neutral Dirac fermions in the $SU(3)$ triplet together with the SM lepton doublet, i.e. the electron and the electron neutrino.  

In order to stabilize the dark matter particle, a matter-parity symmetry is evoked. The dark matter phenomenology is then governed by gauge principles with predictive signals at direct detection and collider experiments. As for direct direction, the main signature stems from spin-independent  WIMP-nucleon scattering at XENON detectors mediated by a neutral vector boson. On the collider side, LHC searches for dilepton events offer a powerful probe significantly cutting into the parameter space of the model \cite{Alves:2016fqe}. The DM relic density is determined by annihilation process into SM states mediated by scalar and pseudoscalar states belonging to the extended Higgs sector as well as by a new $Z'$ gauge boson (it will be shown below that, instead, annihilation processes into mediator final states cannot contribute to the DM relic density). 

We emphasize that our work is different from previous studies that connected lepton flavor violation \cite{Masiero:2004vk,Sierra:2008wj,Kumar:2009sf,Batell:2011tc,Kamenik:2011nb,Carone:2011ur,Masina:2012hg,Lee:2014rba,Baek:2015fma,Heurtier:2016iac,Chekkal:2017eka,Das:2017ski} for the following reasons:

{\bf (i)} We discuss LFV in a different UV complete model.

{\bf (i)} Lepton flavor violation is mediated at tree-level by a WIMP. 

{\bf (iii)} We put our results into perspective with direct detection, collider, and lepton flavor violation experiments, to show that our model is amenable to existing constraints and can be nicely tested by the next generation of experiments.

The paper is organized as follows: We first describe the model, with focus on the relevant ingredients for our reasoning. Later we present the collider bounds. Further, we provide a dark matter phenomenology and put our findings into perspective. Lately, we discuss LFV and draw our conclusions.  

\section{Model}
\label{sec1}

Models that embed the SM weak gauge group into a  $SU(3)$ are growing in interest since they feature rich phenomenology and are able to answer interesting theoretical questions \cite{Pisano:1991ee,Foot:1992rh,Foot:1994ym,Hoang:1995vq,Sanchez:2001ua,Diaz:2003dk,Dias:2005yh,Dong:2006mg,Hernandez:2013hea,Dong:2013wca,Doff:2015nru,Huong:2015dwa,Reig:2016vtf,Pires:2016vek,Reig:2016tuk,Deppisch:2016jzl,Reig:2017nrz,Hati:2017aez,Ferreira:2017mvp,CarcamoHernandez:2017cwi}. In particular, we will concentrate our efforts on models based on the
$SU(3)_L \otimes U(1)_N$ gauge group. We will assume that the SM leptons are placed in the fundamental representation of $SU(3)_L$ along with heavy neutral fermions. Below we briefly discuss the particle content and key features of the model to facilitate our reasoning.


\subsection*{Fermion Content}

Leptons are arranged in the fundamental presentation of SU(3) and in singlets as follows,
\begin{eqnarray}
f_{aL} & = & \left (
\begin{array}{c}
\nu_{a} \\
e_{a} \\
N_{a}
\end{array}
\right )_L\sim(1\,,\,3\,,\,-1/3)\nonumber\\
     &e_{aR}& \sim(1,1,-1)\,,\, N_{aR}\,\sim(1,1,0),
\label{eq:L}
\end{eqnarray}
where $a=1,2,3$ runs over the three generations, and $N_{a(L,R)}$ are new heavy fermions added to the SM particle content, identified as dark matter candidates as we shall see further below. The quantum numbers in Eq.\ref{eq:L} refer to the  $SU(3)_c \otimes SU(3)_L\otimes U(1)_N$ group. For example, the leptonic triplets are color singlets, $SU(3)_L$ triplets and have hypercharge $N=-1/3$, thus, transform as $\sim (1\,,\,3\,,\,-1/3)$.

In a similar vein, quarks are placed in triplets under $SU(3)_L$. In order to cancel the triangle anomalies, two generations are in the adjunct representation of SU(3). Overall, they are given by $(i=1,2)$

\begin{eqnarray}
&&Q_{iL} = \left (
\begin{array}{c}
d_{i} \\
-u_{i} \\
D_{i}
\end{array}
\right )_L\sim(3\,,\,\bar{3}\,,\,0)\,, \nonumber \\
&&
u_{iR}\,\sim(3,1,2/3),\,\,\,
\,\,d_{iR}\,\sim(3,1,-1/3)\,,\,\,\,\, D_{iR}\,\sim(3,1,-1/3),\nonumber \\
&&Q_{3L} = \left (
\begin{array}{c}
u_{3} \\
d_{3} \\
U_{3}
\end{array}
\right )_L\sim(3\,,\,3\,,\,1/3)\,, \nonumber \\
&&
u_{3R}\,\sim(3,1,2/3),
\,\,d_{3R}\,\sim(3,1,-1/3)\,,\,U_{3R}\,\sim(3,1,2/3)
\label{quarks} 
\end{eqnarray}
where $U_3$, $D_1$, and $D_2$ are exotic heavy quarks that have the same electric charge as the ordinary quarks. The electric charge of $U_3$ is $2/3$, whereas of $D_1/D_2$ is $-1/3$, These exotic quarks have a baryon number of $1/3$ but also feature a unit of lepton number. Such particles are referred to in the literature as leptoquarks.

\subsection*{Scalar Content}
\label{scalarcontent}

In order to successfully break $ SU(3)_L\otimes U(1)_N$ into $U(1)_{QED}$, reproducing the SM spectrum, one needs at least three scalar triplets \footnote{See \cite{Ferreira:2011hm,Phong:2013cfa} for a two triplet version of this model, which has been excluded \cite{Dong:2014bha}.} namely,
 
\begin{eqnarray}
\eta = \left (
\begin{array}{c}
\eta^0 \\
\eta^- \\
\eta^{\prime 0}
\end{array}
\right ),\,\rho = \left (
\begin{array}{c}
\rho^+ \\
\rho^0 \\
\rho^{\prime +}
\end{array}
\right ),\,
\chi = \left (
\begin{array}{c}
\chi^0 \\
\chi^{-} \\
\chi^{\prime 0}
\end{array}
\right )\,.
\label{scalarcont} 
\end{eqnarray}

leading to the following potential:

\begin{eqnarray} V(\eta,\rho,\chi)&=&\mu_\chi^2 \chi^2 +\mu_\eta^2\eta^2
+\mu_\rho^2\rho^2+\lambda_1\chi^4 +\lambda_2\eta^4
+\lambda_3\rho^4+ \nonumber \\
&&\lambda_4(\chi^{\dagger}\chi)(\eta^{\dagger}\eta)
+\lambda_5(\chi^{\dagger}\chi)(\rho^{\dagger}\rho)+\lambda_6
(\eta^{\dagger}\eta)(\rho^{\dagger}\rho)+ \nonumber \\
&&\lambda_7(\chi^{\dagger}\eta)(\eta^{\dagger}\chi)
+\lambda_8(\chi^{\dagger}\rho)(\rho^{\dagger}\chi)+\lambda_9
(\eta^{\dagger}\rho)(\rho^{\dagger}\eta) \nonumber \\
&&-\frac{f}{\sqrt{2}}\epsilon^{ijk}\eta_i \rho_j \chi_k +\mbox{H.c}.
\label{potential}
\end{eqnarray}
with $\eta$ and $\chi$ both transforming as $(1\,,\,3\,,\,-1/3)$
and $\rho$ transforming as $(1\,,\,3\,,\,2/3)$. After the spontaneous symmetry breaking led by the non-zero vacuum expectation values of the neutral scalars $\eta^0, \rho^0\, \chi^{\prime 0}$ $\rightarrow v,v, v_{\chi^\prime}$ (where we assumed the first two vevs are equal for simplicity), we are left with two $3\times 3$ mass matrices, one for the CP-even scalars and other for the CP-odd ones. The CP-even matrix leads to three mass-eigenstates,
$S_1,S_2$ and $H$, with
\begin{eqnarray}
M_{S_1}^2 = \frac{1}{2}( v^2/2 + 4 v_{\chi^\prime}^2\lambda_1) \nonumber\\
M_{S_2}^2 = \frac{1}{2}(  v_{\chi^\prime}^2 + 2 v^2(\lambda_2+\lambda_3-\lambda_6) )\nonumber\\
M_H^2= v^2 (\lambda_2+\lambda_3+\lambda_6),
\label{Eq:massscalar}
\end{eqnarray}where H is identified as the Higgs boson with $M_H=125$~GeV. Thus, we have $\sqrt{2} v=246\,\mbox{GeV}$, while $v_{\chi^\prime}$ is a free parameter, but assumed to be much larger than $1$~TeV for consistency. 

The CP-odd mass matrix generates one massive pseudo-scalar, namely,

\begin{equation}
M_{P_1}^2 = \frac{1}{2}( v_{\chi^\prime}^2 + \frac{v^2}{2}),\end{equation}and the mixing between $\chi^0$ and $\eta^{0\prime}$ yields $\phi_1$ where,

\begin{equation} 
M_{\phi_1}^2= \frac{(\lambda_7+1/2)}{2}( v^2+ v_{\chi^\prime}^2).
\end{equation}

Moreover, the charged scalar fields form two separate mass matrices; one involving the fields $\chi^-$ and $\rho^{\prime-}$, and the other containing $\eta^-$ and $\rho^-$. They lead to two physical massive charged scalars,

\begin{eqnarray}
M_{h_1^-} =\frac{\lambda_8+1/2}{2} (v_{\chi^\prime}^2 + v^2),\nonumber\\
M_{h_2^-} =v_{\chi^\prime}^2/2 + \lambda_9 v^2.
\end{eqnarray}

Five goldstone bosons arise, providing the necessary degrees of freedom to generate a longitudinal component for the massive five extra gauge bosons introduced by extending $SU(2)_L$ to $SU(3)_L$. 

\subsection*{Gauge Bosons}
\label{gaugebosons}

As a direct consequence of the enlarged electroweak gauge group, five extra gauge bosons are predicted in the model. We label them as $Z^{\prime}, W^{'\pm},$ and $U^{0}$ and $U^{0\dagger}$. These bosons have masses proportional to the scale of symmetry breaking of the model, i.e. $v_{\chi^\prime}$.  
The hadronic sector is not important for our discussion, so it is set aside. The relevant neutral and charged currents contain,
\begin{eqnarray}
&&{\cal L}  \supset
-\frac{g}{\sqrt{2}}\left[\bar{N}_{aL}\gamma^\mu \ell_{aL} W^{'+}_\mu+ \bar{\nu}_{aL}\gamma^\mu N_{aL} U^0_\mu \right]\nonumber\\ 
&&-\frac{g}{\sqrt{2}}\left[\bar{U}_{3L} \gamma^\mu d_{3L} W^{'+}_\mu + \bar{u}_{iL} \gamma^\mu D_{iL} W^{'+}_\mu\right]\nonumber\\
 &&-\frac{g}{2 \cos\theta_W}\sum_{f} \left[\bar
f\,  \gamma^\mu\ (g^\prime_V + g^\prime_A \gamma^5)f \, { Z_\mu^\prime} \right]
\label{Eq:currents}
\end{eqnarray}with $\ell =e,\mu,\tau$, $f$ representing all fermions in TABLE I, and $g_{V/A}^\prime$ defined accordingly. Notice that the $W^\prime$ gauge boson is always accompanied by an exotic field, either the dark matter particle, N, or exotic quarks $U_3$ and $D_i$. Consequently, the $W^\prime$ gauge boson cannot be singly produced at the LHC, and for this reason, the most stringent collider bound on this models comes from the $Z^\prime$ resonance production as we discuss later on.

The masses of these gauge bosons are found to be \cite{Profumo:2013sca},

\begin{table}[t]
\begin{footnotesize}
\begin{center}
\begin{tabular}{|c|c|c|}
\hline
Interaction &  $g^\prime_V$ & $g^\prime_A$   \\ 

\hline
$Z^{\prime}\ \bar u u,\bar c c  $ &
$\displaystyle{\frac{3-8\sin^2\theta_W}{{6\sqrt{3-4\sin^2\theta_W}}}}$  & 
$\displaystyle{-\frac{1}{2\sqrt{3-4\sin^2\theta_W}}}$  \\
\hline
$Z^{\prime}\ \bar t t$ & 
$\displaystyle{\frac{3+2\sin^2\theta_W}{{6\sqrt{3-4\sin^2\theta_W}}}}$  & 
$\displaystyle{-\frac{1-2\sin^2\theta_W}{2\sqrt{3-4\sin^2\theta_W}}}$  \\
\hline
$Z^{\prime}\ \bar d d,\bar s s  $ &
$\displaystyle{\frac{3- 2\sin^2\theta_W}{6\sqrt{3-4\sin^2\theta_W}}}$  & 
$\displaystyle{-\frac{{3-6\sin^2\theta_W}}{6\sqrt{3-4\sin^2\theta_W}}}$  \\
\hline
$Z^{\prime}\ \bar b b$ & 
$\displaystyle{\frac{3-4\sin^2\theta_W}{{6\sqrt{3-4\sin^2\theta_W}}}}$  & 
$\displaystyle{-\frac{1}{2\sqrt{3-4\sin^2\theta_W}}}$  \\
\hline
$Z^{\prime}\ \bar \ell \ell $ &
$\displaystyle{\frac{-1+4\sin^2\theta_W}{2\sqrt{3-4\sin^2\theta_W}}}$ &
$\displaystyle{\frac{1}{2\sqrt{3-4\sin^2\theta_W}}}$ \\
\hline
$Z^{\prime} \overline{N_i} N_i $ &
$\displaystyle{\frac{4\sqrt{3-4\sin^2\theta_W}}{9}}$ &
$\displaystyle{-\frac{4\sqrt{3-4\sin^2\theta_W}}{9}}$ \\
\hline
$Z^{\prime}\ \overline{\nu_{\ell}} \nu_{\ell} $ &
$\displaystyle{\frac{\sqrt{3-4\sin^2\theta_W}}{18}}$ &
$\displaystyle{-\frac{\sqrt{3-4\sin^2\theta_W}}{18}}$ \\
\hline
\end{tabular}
\end{center}
\end{footnotesize}
\caption{$Z^\prime$ interactions with fermions, where $g_V^\prime$ and $g_A^\prime$ are the vector and axial-vector couplings in the neutral current in Eq.\ref{Eq:currents} .}
\label{tab2}
\end{table}

\begin{eqnarray}
&& M^2_{W^{\prime\pm}} = M^2_{U^0} = \frac{1}{4}g^2(v_{\chi^\prime}^2+v^2)\nonumber\\
&& M^2_{Z^\prime} = \frac{g^{2}}{4(3-4s_W^2)}[4c^{2}_{W}v_{\chi^\prime}^2 +\frac{v^{2}}{c^{2}_{W}}+\frac{v^{2}(1-2s^{2}_{W})^2}{c^{2}_{W}}],\nonumber\\
\label{massvec}
\end{eqnarray}
where  $s_W^2=1-c_W^2 \approx 0.23$ is the sine of the Weinberg angle squared.

Under the assumption $v_{\chi'} \gg v$ they can be approximately expressed as $M_{Z^\prime} \simeq 0.4 v_{\chi^\prime}$ and $M_{W^\prime} \simeq 0.32 v_{\chi^\prime}$ respectively. Under the same approximation and for $\lambda_1=1$, $M_{P_1} = 0.7 v_{\chi^\prime}$, and  $M_{S_1} \simeq 1.4 v_{\chi^\prime}$. As evidenced by the approximate expressions above, in order to achieve multi-TeV masses for the $Z^\prime$ and $W^\prime$, to comply with collider bounds, we need to assume that the $SU(3)_L$ symmetry is broken at sufficiently high energy scale. We shall see further that having these mass relations handy will help us understand the dark matter phenomenology.

\subsection*{Dark Matter Stability}

In order to ensure that we have a viable dark matter candidate we invoke a matter-symmetry which mimics the supersymmetric standard model reading, $P=(-1)^{3(B-L)+2s}$, where $B$ is the baryon number, $L$ is the lepton number and $s$ is spin of the field as follows,

\begin{eqnarray}
(N_i\,,\,D_1,D_2\,,\,U_3\,,\,\rho^{\prime +}\,,\,\eta^{\prime 0}\,,\,\chi^{0}\,,\,\chi^-\,,\, W^\prime\,,\,U) \rightarrow -1.\nonumber\\
\label{discretesymmetryI}
\end{eqnarray}

The remaining fields transform trivially under this matter-symmetry. Many of the fields above have non-trivial lepton number as shown in \cite{Dong:2014wsa}. As in supersymmetry, the lightest odd-neutral particle is stable and a potential dark matter candidate. That said, it is important to consider the charged current in Eq.\ref{Eq:currents}. For the dark matter fermion to be stable its mass has to be below the $W^\prime$ mass. However, we have seen above that $M_{W^\prime} \simeq 0.32 v_{\chi^\prime}$. Therefore, for a given scale of symmetry breaking, there is an upper limit on the dark matter mass due to stability requirements. If the dark matter is heavier than the $W^\prime$ mass, we have a scenario of decaying dark matter that is tightly constrained by data \cite{Audren:2014bca}. This stability condition reflects into a gray region toward the bottom of FIG.\ref{figMZpMDM}, see below.

\subsection*{Fermion Masses}

All fermions obtain dirac masses generated through the Yukawa lagrangian \cite{Mizukoshi:2010ky}, 
\begin{eqnarray}
&-&{\cal L}^Y =\alpha_{ij} \bar Q_{iL}\chi^* D_{jR} +f_{33} \bar Q_{3L}\chi U_{3R} + g_{ia}\bar Q_{iL}\eta^* d_{aR} \nonumber \\
&&+h_{3a} \bar Q_{3L}\eta u_{aR} +g_{3a}\bar Q_{3L}\rho d_{aR}+h_{ia}\bar Q_{iL}\rho^* u_{aR} \nonumber \\
&&+ G_{ab}\bar f_{aL} \rho e_{bR}+g^{\prime}_{ab}\bar{f}_{aL}\chi N_{bR}+ \mbox{h.c}. 
\label{yukawa}
\end{eqnarray}

The SM model spectrum is successfully reproduced in this way. One important remark that needs to be made concerning Eq.\ref{yukawa} is the fact that the dark matter mass is found to be,

\begin{equation}
M_{N_1}= g^\prime v_{\chi^\prime}/\sqrt{2}
\end{equation}

Since the Yukawa couplings $g^\prime$ are free, these Dirac fermions can in general mix. In this case, the dark matter particle would actually be a composition of three states. Another crucial feature is the origin of lepton flavor violation. Since the off-diagonal couplings can be non-zero, there will be mixing matrices. These mixing matrices will enter into the charged current involving the $W^\prime$ gauge boson in Eq.\ref{Eq:currents} and induce the $\mu \rightarrow e \gamma$ decay. It is also important to highlight that we will never adopt $g^\prime \sim 1$, because if $g^\prime \sim 1$, then $M_{N_1} \sim 0.7 v_{\chi}^{\prime}$, making the dark matter heavier than the $W^\prime$ boson. As we have discussed before, this would lead to an excluded decaying dark matter scenario due to the charged current involving the $W^\prime$ gauge boson.

Having revised the key ingredients of the model, we will present, in the following, the existing and projected collider limits \footnote{The model might be subject to other limits but they are subdominant compared to the collider ones \cite{Hue:2015fbb,Benavides:2015afa,Boucenna:2015zwa,Salazar:2015gxa,Sanchez-Vega:2016dwe,Hue:2017lak,Santos:2017jbv}}.\\
  
\section{Collider Limits}

There have been several collider searches for new gauge bosons at the LHC. Recently, new limits have been derived in the context of 3-3-1 models. Looking into the dilepton data the authors of \cite{Alves:2016fqe} found $M_{Z^\prime} > 4.1$TeV, at $13$~TeV with $36.1fb^{-1}$ of integrated luminosity.  Projections of this limit for 13TeV center-of-energy with $100 (1000) fb^{-1}$ of integrated-luminosity reach $M_{Z^\prime} > 4.9 (6.1)$~TeV \cite{Alves:2016fqe}. These limits are represented as dashed lines in FIG.\ref{figMZpMDM}. Existing bounds on $W^\prime$ gauge bosons based on signal events with a charged lepton plus missing energy are not applicable to our model because our $W^\prime$ cannot be singly produced at the LHC.
Therefore, the most effective collider bounds stem from signal events with dileptons. Notice though, that we can still place a lower mass bound on the $W^\prime$ mass by using the relation $M_{Z^\prime}= 1.25 M_{W^\prime}$ which comes straightforwardly from Eq.\ref{massvec}. 

\begin{figure*}
\includegraphics[scale=0.3]{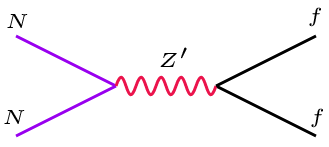}
\includegraphics[scale=0.3]{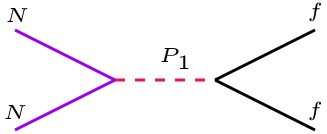}
\includegraphics[scale=0.3]{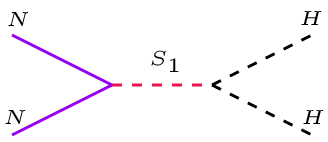}
\includegraphics[scale=0.4]{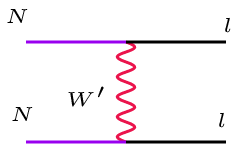}
\includegraphics[scale=0.4]{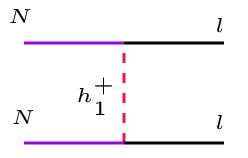}
\caption{Feynman diagrams relevant for the dark matter phenomenology. The first, second and third diagram refers to an s-channel annihilation through a $Z^\prime$, a pseudo-scalar and a scalar, respectively. The fourth and fifth diagrams are t-channel induced via exchange of either a $W^\prime$ or a charged Higgs. The t-channel versions of first and second diagrams are also responsible for DM-nucleon scattering. The scattering rate induced by a pseudo-scalar is momentum suppressed at the fourth power, therefore only the t-channel $Z^\prime$ exchange is important for direct dark matter detection.}
\label{feynmanndia}
\end{figure*}

\section{Dark Matter}
\label{darkmatter}

Dark matter has been addressed in the context of $SU(3)\otimes U(1)_X$ models in many forms and in different models \cite{Filippi:2005mt,Kelso:2013nwa,Dong:2014esa,Dong:2015rka,Martinez:2015wrp,Ferreira:2016uao}. The most important dark matter observables in our model are the relic density and the scattering rate on nucleons. Indirect dark matter detection constraints stemming from gamma-ray observations for instance are mostly relevant for dark matter masses below $100$~GeV or masses far above the TeV scale out of reach of colliders \cite{Acharya:2017ttl,Balazs:2017hxh}. For this reason, they will not be discussed in detail.  

Any of the neutral fermions could be the lightest field, thus the dark matter, but without loss of generality we will assume to be $N_1$. The existence of non-zero mass mixing between the heavy fermions yield negligible impact to our dark matter phenomenology, in agreement with \cite{Profumo:2013sca}. There are several diagrams contributing to the relic density of the fermion, see FIG.\ref{feynmanndia}. The first diagram refers to the s-channel $Z^\prime$ exchange. The second diagram corresponds to s-channel annihilation mediated by the pseudo-scalar $P_1$.  The third originates from the scalar $S_1$ that can also contribute via s-channel annihilation into Higgs pairs. The fourth and fifth diagrams account for the t-channel exchange of either a $W^\prime$ or a charged Higgs. 

The last two have been found to be small compared to the s-channel and t-channel interactions mediated by the $Z^\prime$ gauge boson, in agreement with \cite{Profumo:2013sca}. Having in mind that $g^{'}_{N_1 V} (g^{'}_{N_1 A})$ is the product between the vector (vector-axial) couplings of the $Z^\prime \bar{N_1}N_1$ and $Z^\prime \bar{f}f$ vertices as shown in TABLE I, the relevant annihilation channels provide,

\begin{widetext}
\begin{eqnarray}
\label{equation: annihilation1}
\langle \sigma v_{N_1} \rangle \left( N_1 \bar{N_1} \to Z^\prime  \to f \bar{f} \right) &\simeq \frac{n_c \sqrt{1-\frac{m_f^2}{\mchi^2}}}{2 \pi \mzp^4 \left( 4 \mchi^2 - \mzp^2 \right)^2} \bigg\{ g_{fA}^{'\,2} \Big[ 2 g_{N_1V}^{'\,2} \mzp^4 \left(\mchi^2 - m_f^2 \right) + g_{N_1 A}^{'\,2} m_f^2 \left( 4 \mchi^2 - \mzp^2\right)^2 \Big]
\nonumber \\
& + g_{N_1 V}^{'\,2} g_{f V}^{\,2} \mzp^4 \left( 2 \mchi^2 + m_f^2 \right) \bigg\}
~,
\end{eqnarray}
\end{widetext}

where $v_{N_1}$ is the relative velocity of the annihilating DM pair, $n_c=1$ for leptons, whereas $n_c=3$ for quarks. An important feature of this model is that the DM particle cannot be heavier than the  $Z^\prime$ boson due to stability requirements, as discussed previously. Therefore, there is no annihilation into a pair of $Z^\prime$ gauge bosons as occurs often in the case of simplified dark matter models. \\

Another important annihilation process is the s-channel annihilation via a pseudo-scalar that yields,

\begin{widetext}
\begin{eqnarray}
& \langle \sigma v_{N_1} \rangle \left( N)1 \bar{N}_1 \to P_1  \to f \bar{f} \right) \simeq \frac{m_f^2 |\lambda_N|^2 M_{N_1} \sqrt{M_{N_1}^2-m_f^2}}{2 \pi v^2  \left(M_{P_1}^2-4 M_{N_1}^2\right)^2}
\end{eqnarray}
\end{widetext}

where $\lambda_N= M_{N_1}\, v/ (\sqrt{2}\, v_{\chi^\prime})$. \\

\begin{figure}
\includegraphics[width=\columnwidth]{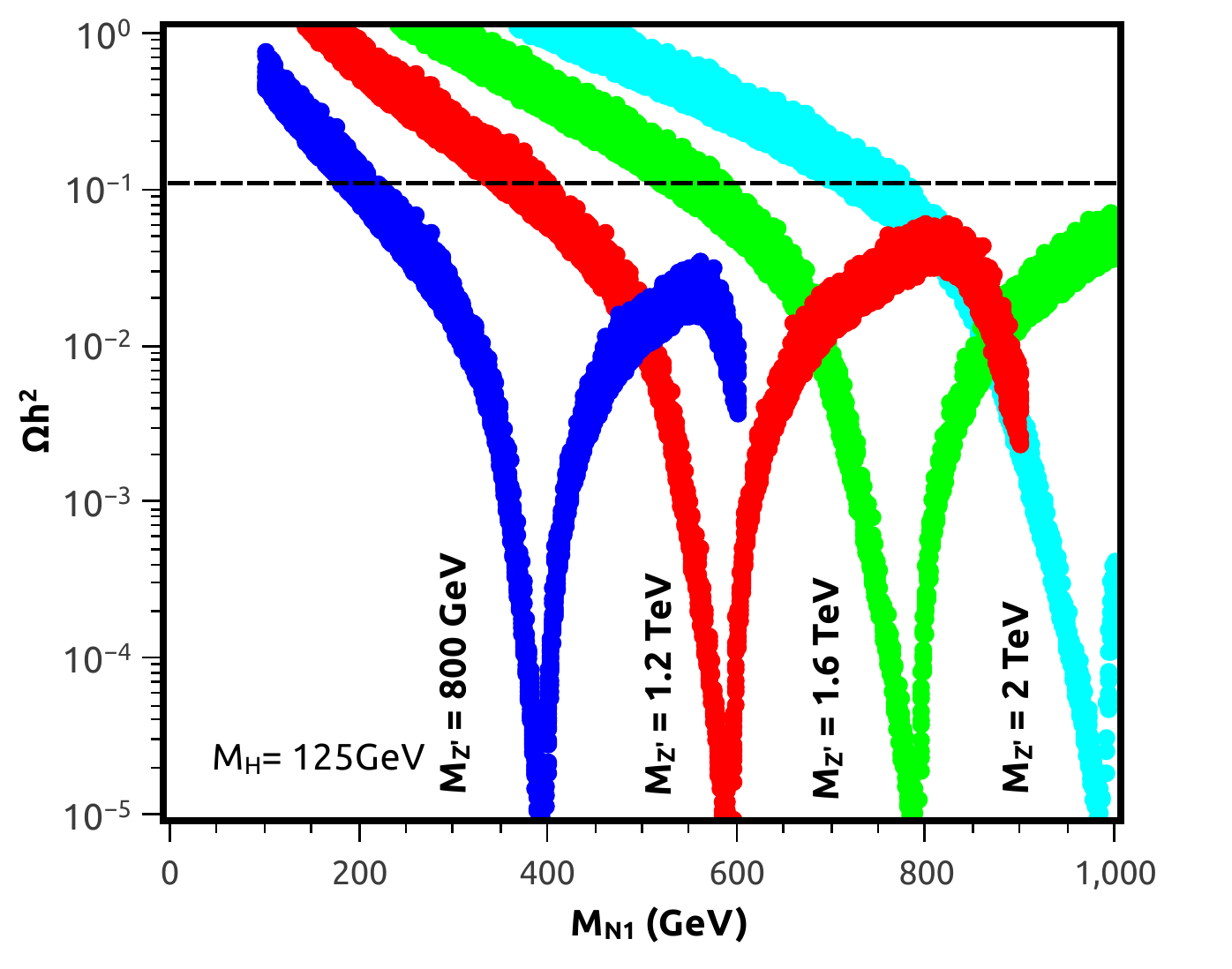}
\caption{Relic density for the Dirac dark matter fermion $N_1$, allowing for co-annihilation with nearly degenerate fermions $N_2$ and $N_3$ for different $Z^\prime$ masses. The impact of co-annihilation is very mild it just changes the thickness of the curves.}
\label{figrelicdens}
\end{figure}

Similarly to what happens in the $Z^\prime$ mediated case, the annihilation via pseudo-scalar features a resonance when $M_{N_1} \sim M_{P_1}/2$. These two annihilation modes drive the relic density as can be observed in FIG.\ref{figrelicdens}. \\

Since we adopted  $\lambda_1$ of O(1) and thus $M_{S_1} \sim 1.4 v_{\chi}$ the third diagram in FIG.\ref{feynmanndia} is not so relevant to our findings.  Since current collider bounds imply that $M_{Z^\prime} > 4.1$~TeV, we get  $v_{\chi^\prime} \gtrsim 10$~TeV, which implies that $M_{S_1} > 14$~TeV, dwindling its relevance.  If we tune down the coupling $\lambda_1$ to bring down the $S_1$ mass to the TeV scale, still we have checked that the impact on the relic dark matter density is very mild. Anyways, we emphasize that we included all processes in the numerical calculations with $\lambda_1 \sim 1$ using the Micromegas package \cite{Belanger:2006is,Belanger:2008sj}. \\

Moreover, the fourth and fifth diagrams that represent t-channel annihilations into charged leptons, possibly even including annihilations into final states that violate lepton flavor, are also subdominant \cite{Profumo:2013sca} \footnote{These lepton flavor violating annihilating channels also arise in other frameworks \cite{Vicente:2014wga}}. We emphasize that in our model dark matter plays a key role in LFV observables but LFV observables do not set the dark matter relic density. However, the dark matter relic and LFV observables can still be tied to each other because everything is ruled by the gauge symmetry, making our model predictive.\\

In summary, with these annihilation cross sections at hand one can compute the dark matter relic abundance using,

\begin{equation}
\Omega_N h^2 \approx 0.1 \frac{x_F}{20} \frac{80}{g_{\star}} \left( \frac{3\times 10^{-26} cm^3 s^{-1}}{\langle \sigma v \rangle} \right),
\label{Eq:abundance}
\end{equation}where $x_F \sim 20-30$, and $g_{\star} \sim 80$.\\

From Eq.\ref{Eq:abundance} we can see that the larger the annihilation cross section the smaller the relic density, and for annihilation cross sections around $10^{-26} cm^3/s$ one can obtain the correct relic density.  Since Eq.\ref{Eq:abundance} provides just an approximation, we in fact performed our calculation numerically within 
the Micromegas package \cite{Belanger:2006is,Belanger:2008sj}.\\

\begin{figure}
\includegraphics[width=\columnwidth]{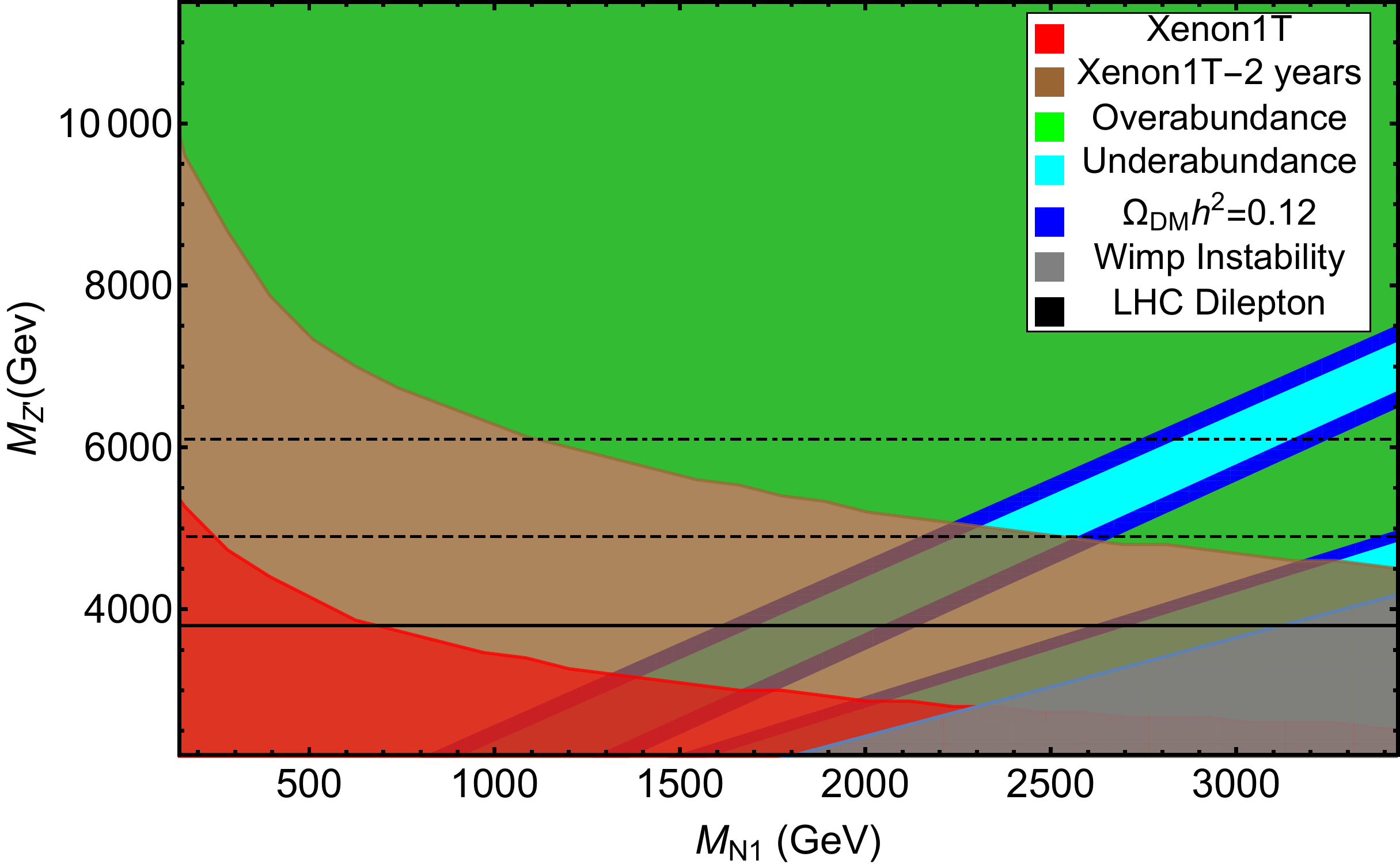}
\caption{A summary plot that includes the dark matter relic density (blue), direct detection constraints (shaded red and brown and collider limits (horizontal black lines), in the $M_{Z^\prime}, M_{N1}$ mass plane. $N_1$ is the lightest neutral fermion in the leptonic triplet of $SU(3)_L$. Current and projected limits are included in the figure. See text for detail.}
\label{figMZpMDM}
\end{figure}

Our results are summarized in FIG.\ref{figrelicdens}, where one can inspect that the $Z^\prime$ resonance is quite visible. To understand the physics behind the curves in FIG.\ref{figrelicdens} let's focus on the result for $M_{Z^\prime}=800$~GeV~\footnote{Notice that in FIG.\ref{figrelicdens} we have not considered bounds from collider searches, hence taking relatively low values of $M_{Z^\prime}$. The figure should be just regarded as an illustration of the behavior of the DM relic density. We remark, nevertheless, that the shape of the curves is not affected by the value of $M_{Z'}$. Higher values would just shift the curves to higher dark matter masses.} for instance. The depth of the blue curve is driven the $Z^\prime$ decay width which governs the dark matter annihilation cross section at the resonance, i.e. when $M_{N_1} \sim M_{Z^\prime}/2$. As the dark matter mass increases we go away from the $Z^\prime$ resonance, and for this reason, the dark matter annihilation cross section decreases as the dark matter abundance increases.  However, at some point the annihilation via the pseudo-scalar kicks in, in particular  when $M_{N_1} \sim M_{P_1}/2$ which occurs for $M_{N_1} \sim 700$~GeV, since $M_{Z^\prime}=800$~GeV , $v_{\chi^\prime}=2$~TeV and $M_{P_1} \sim 1.4$~TeV. For this reason, we can see the blue curve turning down again. We emphasize that the dark matter annihilation into a pair of mediators is not present in this model, scenario known as secluded dark matter \cite{Profumo:2017obk}, because the dark matter mass cannot be larger than the $W^\prime$ mass, i.e. $0.8 M_{Z^\prime}$, otherwise it would decay. The scenarios shown in FIG.\ref{figrelicdens} are for $Z^\prime$ masses below $2$~TeV, which have been excluded by the LHC, but they serve their purpose which is to easily capture the physics behind the relic density. In FIGs.\ref{figMZpMDM}-\ref{Fig.LFVplot} we extend the relic density curves to heavier dark matter masses and put in context with other observables.\\

Regarding the dark matter-nucleon scattering rate, it is mostly due to spin-independent interaction from t-channel $Z^\prime$ exchange. The pseudoscalar leads to a unobservable scattering cross section for dark matter masses above $100$~GeV \cite{Arcadi:2017wqi}, thus the direct detection of dark matter is ruled by the $Z^\prime$ boson. The corresponding spin-independent scattering cross section features the well-known $A^2$ (atomic mass) enhancement. For this reason, experiments with heavy targets such as XENON1T, LUX and PANDA-X provide the most restrictive limits in the literature \cite{Akerib:2016lao,Aprile:2016swn,Fu:2016ega,Tan:2016zwf,Aprile:2017iyp,Akerib:2017kat}. In summary, the dark matter-nucleon scattering cross section is parametrized as, 

\begin{equation}
\label{equation: SI}
\sigma^{SI} \approx \frac{\mu^2_{N n}}{\pi} \Big[ \frac{Z f_p + (A-Z) f_n}{A} \Big]^2,
\end{equation}where, 

\begin{equation}
f_p \equiv \frac{1}{M_{Z^\prime}^2 } \left( 2 g^{'}_{u V} + g^{'}_{d V} \right),
\end{equation}and

\begin{equation}
f_n \equiv \frac{1}{M_{Z^\prime}^2} \left( g^{'}_{u V} + 2 g^{'}_{d V} \right),
\end{equation}where  $\mu_{\chi n}$ is the dark matter-nucleon reduced mass, $g_{uV}$ and $g_{dV}$ are the (vectorial) $N$ couplings to up- and down-quarks which are simply the product of the $Z^\prime-q-q$ with $N-N-Z^\prime$ couplings as given in Table I, where $q=u,d$. $Z$ and $A$ are the atomic number and atomic mass of the target nucleus, respectively.

We now combine all results into FIG.\ref{figMZpMDM}. The parameter space that yields the right relic density is shown in dark blue. The cyan region leads to $\Omega _{N_1} h^2 < 0.1$, i.e. underabundance. The region in green produces  $\Omega _{N_1} h^2 > 0.1$. We highlight that we have set aside any non-standard cosmology effects that could potentially change the region of parameter space that yields the correct relic density \cite{Gelmini:2006mr,Gelmini:2006pw,Arcadi:2011ev,Baer:2014eja}. 

In FIG.\ref{figMZpMDM} the collider limits on the $Z^\prime$ mass are delimited by horizontal black lines. From bottom to top, these limits represent current and projected exclusion for $100$ and $1000 fb^{-1}$ of integrated luminosity. When the $M_{Z^\prime} > 2 M_{N_1}$ the invisible decay into dark matter opens, however, the branching ratio is rather small and for this reason the collider limits do not weaken when $M_{Z^\prime} > 2 M_{N_1}$ in FIG.\ref{figMZpMDM}. (See  \cite{Arcadi:2017kky} for discussion on this topic.)

The direct detection limits scale with $1/M_{Z^\prime}^4$, as can be seen in Eq.15-17. This scaling can be observed in FIG.\ref{figMZpMDM}. There we show in red the current exclusion bounds on the spin-independent scattering cross section from XENON1T \cite{Aprile:2017iyp}, and in brown the projected limits from XENON1T with 2-years exposure \cite{Aprile:2015uzo}. The complementarity between LHC and direct dark matter detection is visible, constituting a strong case for these independent searches and the importance of further data taking.

The region in gray delimits the instability region of the dark matter particle. Both the $Z^\prime$ and $W^\prime$ masses are directly connected to the scale of symmetry breaking of the model as we discussed previously. Remembering that the charge current term, $\bar{N_1}\gamma^\mu e W^+_\mu$, $W^\prime$ prohibits the dark matter particle to be heavier than the $W^\prime$, for a given $W^\prime$ mass, where $M_{W^\prime} =0.8 M_{Z^\prime}$ mass, there is maximum allowed value for the dark matter mass in which the dark matter remains stable. This stability requirement gives rise to the gray region in FIG.\ref{figMZpMDM}, region which the dark matter particle is unstable, in order words, $M_{N_1} > 0.8 M_Z^{\prime}$.

In summary, taking into account, relic density, direct dark matter detection and collider constraints, we can conclude that our model furnishes a viable dark matter model in agreement with current and projected constraints. We will put all these observables now into perspective with LFV observables in the next section.

\section{Lepton Flavor Violation}

\begin{figure}[!h]
\includegraphics[width=0.6\columnwidth]{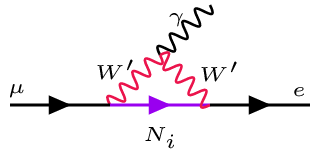}
\caption{Diagram that contributes to the $\mu \rightarrow e\gamma$ decay. The mixing between the dark matter fermions $N_i$ is responsible to the LFV signal. 
Since we adopted $N_1$ to be the lightest fermion, we take only the leading order terms involving $N_1$, and set aside the other similar diagrams involving the heavier fermions.}
\label{figdiagramLFV}
\end{figure}

Lepton flavor violation is a smoking gun signature for physics beyond the SM \cite{Lindner:2016bgg,Calibbi:2017uvl}.  The observation of neutrino oscillations has solidly shown us that lepton flavor is not an exact symmetry in nature. Thus it is plausible to assume that such violation also occurs between charged leptons. In particular, processes such as $\mu \rightarrow e\gamma$, $\mu \rightarrow 3e$ and $\mu-e$ conversion are great laboratories for new physics \cite{Lindner:2016bgg}. Currently, we dispose of very strong limits on the branching ratio of these LFV muon decay modes as well as on the $\mu-e$ conversion rate.  In our model, these LFV muon decays occur via a diagram that involves the $W^\prime$ and the dark matter particles, $N_i$ as shown in FIG.\ref{figdiagramLFV}. 

We highlight that we adopted $N_1$ to be the lightest fermion throughout, therefore the similar diagrams involving the other heavy fermions are subdominant. We conservatively, take only the leading order terms involving $N_1$, and set aside the other similar diagrams involving the heavier fermions. Letting $g^{N_1e}= g/(2\sqrt{2}) U^{N_1 e}$ and $g^{N_1\mu}= g/(2\sqrt{2}) U^{N_1 \mu}$, we find that,

\begin{equation}
Br(\mu \rightarrow e\gamma) =\frac{3 (4\pi)^3 \alpha_{em}}{4 G_F^2}(|A^M_{e\mu}|^2 + |A^{E}_{e\mu}|^2 ) 
\label{eqLFV}
\end{equation}with, 

\begin{eqnarray}
A^M_{e\mu}&=&\frac{-1}{(4\pi)^2} (g^{N_1e \ast}g^{N_1\mu} I_{N_1,3}^{++} +g^{N_1 e \ast}g^{N_i\mu} I_{N_1,3}^{+-})\nonumber\\
A^E_{e\mu}&=&\frac{i}{(4\pi)^2}(g^{N_1 e \ast}g^{N_1\mu} I_{N_1,3}^{-+} +g^{N_1 e \ast}g^{N_1\mu} I_{N_1,3}^{--}),\nonumber\\
\end{eqnarray}where the lengthy integral functions $I_{N_1,3}$ are given in \cite{Lindner:2016bgg}. It is useful to have an analytical approximation corresponding to the case $M_{N_1} \equiv M_{W^\prime}$ to ease the understanding of the LFV signal region,

\begin{equation}
Br (\mu \rightarrow e\gamma)=1.6 \times \left(\frac{1\, {\rm TeV}}{ M_{W^\prime}}\right)^4 |g^{N_1 e \ast} g^{N_1 \mu}|^2.
\label{Eq.LFVapprox}
\end{equation} 

\begin{figure}
\includegraphics[width=\columnwidth]{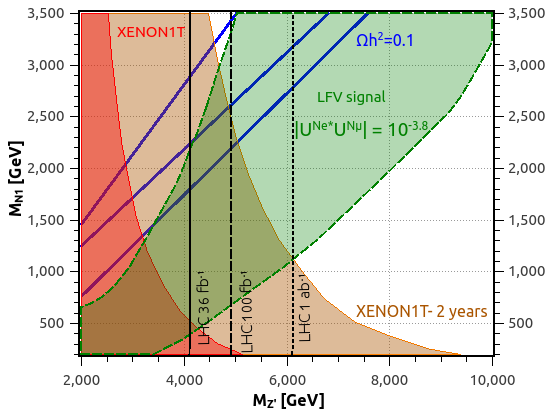}
\caption{Signal region for lepton flavor violation in green, overlaid with current and projected bounds stemming from direct dark matter detection and collider. The dark matter relic density curve is in blue, the collider bounds are represented by vertical black lines, and the current and projected direct detection bounds by red and brown regions respectively.}
\label{Fig.LFVplot}
\end{figure}

However, for our analysis we have employed a full numerical determination of these integrals. We show, in FIG.\ref{Fig.LFVplot}, the signal region for the $\mu \rightarrow e\gamma$ decay, defined as the parameter space that induces a branching ratio below the current bound,  $Br (\mu \rightarrow e\gamma) < 4.2 \times 10^{-13}$ but above the projected sensitivity $Br (\mu \rightarrow e\gamma) \approx 10^{-14}$. The LFV signal depends on the $W^\prime$, but we plot the results in terms of the $Z^\prime$ and the dark matter masses, by using $M_{W^\prime}= 0.8 M_{Z^\prime}$ as explained earlier. We emphasize that we solved Eq.\ref{eqLFV} numerically since Eq.\ref{Eq.LFVapprox} is valid only for a special case. The solution gives rise to the green region shown in FIG.\ref{Fig.LFVplot} which delimits the parameter in which a signal in the $\mu \rightarrow e\gamma$ decay can take place.

There are three free parameters governing this signal region, namely the dark matter mass, the $Z^\prime$ mass (related to the mass of the $W^\prime$ by a constant factor) and the product of the mixing matrices. In order to report a contour in the $(M_{Z'},M_{N_1})$ bidimensional plane, we needed then to fix the value of $g^{N_1 e \ast} g^{N_1 \mu}$. Given that this product appears in the numerator of $Br (\mu \rightarrow e\gamma)$, it can be easily argued that large values of this product would imply into larger $Z^\prime$ masses. Therefore, this would result in a shift of the green region in FIG.\ref{Fig.LFVplot} toward larger $Z^\prime$ masses, eventually putting it outside the isocontours (blue solid lines) of the correct DM relic density. Too small values of $g^{N_1 e \ast} g^{N_1 \mu}$ would be, similarly not be interesting, since they would require low values of the $M_{Z'}$ excluded by LHC and direct DM searches. Keeping the dark matter mass not far from a few TeV as displayed in FIG.\ref{Fig.LFVplot}, and having this logic in mind, we found that $|U^{eN_1\ast} U^{\mu N_1}| = 10^{-3.8}$ yields the largest LFV signals while still reproducing the correct dark matter relic density via the $Z^\prime$ and pseudoscalar resonances. Therefore, keeping the dark matter mass not far from the TeV, our model has an approximate prediction for the largest LFV signal one can have while reproducing the correct dark matter relic density. 
 
It is important to note that the mixing between the dark matter fermions is paramount to generate the LFV signal, conversely it is nearly irrelevant to the dark matter phenomenology. Therefore, the existence of a viable dark matter candidate does not depend on the existence of a LFV signal, but the LFV signal do rely on the dark matter properties, since it mediates the processes. Moreover, since the gauge bosons that dictate the dark matter phenomenology and the LFV signal are related in mass as ruled by the gauge symmetry, one can tie the dark matter relic density and direct detection scattering to the collider and LFV observables. In other words, the fact that we have gauge principles ruling the phenomenology makes our reasoning valid and appealing.

A similar process to FIG.\ref{figdiagramLFV} with an off-shell photon induces the $\mu \to 3e$ decay, with $Br(\mu \rightarrow eee)\sim 1/160 Br(\mu \rightarrow e\gamma)$ \cite{Lindner:2016bgg}. As for $\mu-e$ conversion, the conversion rate can be approximated to be  $CR (\mu-e) \sim 1/200 \, Br(\mu \rightarrow e\gamma)$ \cite{Lindner:2016bgg}. Therefore, one could also draw signal regions for both observables. Having in mind that the current (projected) bound on $(\mu \rightarrow eee)$ reads $10^{-12} (10^{-16})$ and $Br(\mu \rightarrow eee)\sim 1/160 Br(\mu \rightarrow e\gamma)$ it is clear that signal region for the $\mu \rightarrow eee$ decay is basically inside the green region in FIG.\ref{Fig.LFVplot}. Hence, no need to draw a new signal region for it.
Concerning $\mu-e$ conversion the situation is expected to change in the next generation of experiments.  Despite having also a smaller rate, the $\mu-e$ conversion rate is currently limited to be smaller than $6.1\times 10^{-13}$, with a projected bound of $10^{-16}-10^{-18}$ \cite{Lindner:2016bgg,Calibbi:2017uvl}. Considering the conservative $10^{-16}$ value, again the signal region for $\mu-e$ conversion nearly overlaps with the one from $\mu \rightarrow e\gamma$. If we had adopted the $10^{-18}$ value as future sensitivity, the green region on FIG.\ref{Fig.LFVplot} would have extended to lower values in $M_{N_1}$ which is not very interesting because for such values the correct relic density is not achieved. In summary, the green region stands for an approximate signal region for all these three LFV observables $\mu \rightarrow e\gamma$,  $\mu \rightarrow eee$, and $\mu-e$ conversion.  

Besides the results from the LFV observables aforementioned, in FIG.\ref{Fig.LFVplot} we overlay the relic density curves (solid blue lines), the current (projected) bounds from direct detection from XENON1T with 34 days (2 years) live-time exposure with red and brown hashed regions, and finally the current and projected collider bounds as dotted vertical black lines.
 
Notice that for $M_{Z^\prime} \sim 5$~TeV and $M_{N_1} \sim 3.5$~TeV, one can evade collider and direct detection bounds while yielding a sizable LFV signal. If we keep the $Z^\prime$ mass constant and lower the dark matter mass to $2.5$~TeV, we lie precisely on the top of the XENON1T-2 years sensitivity, while still reproducing the correct relic density. Notice that the correct relic density requirement fixes the $Z^\prime$ mass since the model needs to live near the $Z^\prime$ resonance. The $Z^\prime$ masses required in this case is within reach of LHC probes. Therefore, if XENON1T experiment starts seeing excess events consistent with $\sim$ TeV dark matter which can be later confirmed by XENONnT \cite{Aprile:2015uzo}, our model offers clear predictions for the LFV and collider observables which can be fully probed in the next generation of experiments. In summary, our model is predictive and falsifiable.   

Different benchmark models can be straightforwardly chosen using FIG.\ref{Fig.LFVplot} but our messages are very clear:\\

{\it Since we have a UV complete model at hand with predictive signatures at collider, direct detection and LFV experiments one can potentially discriminate our model from others in the literature.}

{\it One can successfully accommodate a dirac fermion dark matter and signal in LFV observables in agreement with direct detection and collider experiments. }

\section{Conclusions}

We have discussed a predictive and falsifiable UV complete model that hosts a dark matter candidate that mediates lepton flavor violation signals such as $\mu \rightarrow e\gamma$, $\mu \rightarrow eee$ and $\mu-e$ conversion. The model features heavy $Z^\prime$ and $W^\prime$ gauge bosons which independently play a role in the dark matter relic density and lepton flavor violation observables. Since our model is ruled by gauge principles, the masses of these gauge bosons are related, $M_{Z^\prime} =1.25 M_{W^\prime}$, for this reason one can exploit the correlation between these observables. 

We find that our model can successfully simultaneously accommodate dark matter and LFV signals in within reach of upcoming experiments. Moreover, if the dark matter mass is not far from the TeV scale, our model offers an approximate prediction for the maximum LFV signal one could induce while reproducing the correct dark matter relic density. 

\section*{Acknowledgements}

The authors thank Werner Rodejohann, Manfred Lindner, Clarissa Siqueira, Alex Dias, Alexandre Alves, Carlos Pires, Paulo Rodrigues, Pedro Pasquini, Orlando Peres and Pedro Holanda for useful discussions. CS is grateful to the Conselho Nacional de Desenvolvimento Cient\'ifico e Tecnol\'ogico (CNPq), CF and MG thank the Funda\c c\~ao de Amparo \`a Pesquisa do Estado de S\~ao Paulo (FAPESP) for the financial support. FSQ acknowledges support from MEC and ICTP-SAIFR FAPESP grant 2016/01343-7.

\bibliography{darkmatter}

\end{document}